\documentclass[11pt,final,journal,twocolumn]{IEEEtran}
\usepackage[T1]{fontenc}
\usepackage[utf8]{inputenc}
\usepackage{textcomp}
\usepackage{graphicx}
\usepackage{ae,aeguill}
\usepackage{float} 
\usepackage{amsfonts}
\usepackage{sistyle}
\newcommand{\sinc}{{\rm sinc}}
\usepackage{bm}

\begin{document} 
\title{Lorentz-Force Hydrophone Characterization}
\author{
Pol Grasland-Mongrain, Jean-Martial Mari, Bruno Gilles, Adrien Poizat, Jean-Yves Chapelon and Cyril Lafon
\thanks{Inserm, U1032, LabTau, Lyon, F-69003, France ; Universit\'e de Lyon, Lyon, F-69003, France} 
}

\maketitle

\begin{abstract}
A Lorentz-force hydrophone consists of a thin wire placed inside a magnetic field. When under the influence of an ultrasound pulse, the wire vibrates and an electrical signal is induced by the Lorentz force that is proportional to the pulse amplitude. In this study a compact prototype of such a hydrophone is introduced and characterized, and the hydrodynamic model previously developed is refined. It is shown that the wire tension has a negligible effect on the measurement of pressure. The frequency response of the hydrophone reaches 1 MHz for wires with a diameter ranging between 70 and 400 \micro m. The hydrophone exhibits a directional response such that the signal amplitude differs by less than 3dB as the angle of the incident ultrasound pulse varies from -20$^o$  and +20$^o$. The linearity of the measured signal is confirmed across the 50 kPa to 10 MPa pressure range, and an excellent resistance to cavitation is observed. This hydrophone is of interest for high pressure ultrasound measurements including High Intensity Focused Ultrasound (HIFU) and ultrasonic measurements in difficult environments.
\end{abstract}


\section{Introduction}
The ultrasound beams used in medical applications must be characterized precisely, especially in the case of therapeutic ultrasound \cite{preston1991} to ensure the safety of the patient and the success of the treatment. This is particularly the case with High Intensity Focused Ultrasound (HIFU), where an ultrasound beam is used to thermally ablate tissues in various organs like liver, bladder, kidney, prostate, breast and brain \cite{hill1995high},\cite{ter2000intervention}, and where any uncertainty on the properties of the ultrasound beam could have a dramatic impact on the efficiency of the treatment, as well as dangerous side effects \cite{shaw2008buoyancy}.

Piezoelectric and piezoceramic hydrophones of different shapes (needle, lipstick or membrane designs) are currently the gold standards for the characterization of ultrasonic transducers \cite{harris2000}. Those hydrophones have a broad frequency response, a wide linear output over pressure, stable properties over time and a negligible impact on the fields they measure \cite{retat2011} \cite{smith1989hydrophones}. However, these hydrophones are fragile and high pressure beams can irreversibly damage them and rapidly impair their ability to function properly. Fiber-optic hydrophones are much less fragile, but demand an expensive setup and have a lower sensitivity, with a typical minimum measurable pressure of about 100 kPa \cite{klann2005}, \cite{staudenraus1993fibre}. Better sensitivities are achieved when a Fabry-P\'erot cavity is created at the tip of the optical fiber \cite{cox2007}, but at the cost of a greater fragility. Schlieren imaging systems \cite{korpel1987}, \cite{newman1973observations} are also used to map ultrasound beams, where the variations of refractive index induced by the pressure pulse are capture optically. In principle it allows quantitative measurements \cite{hanafy1991}, however the linearity of the refractive index with pressure can only be assumed at low pressure \cite{pitts2000three}. Furthermore difficulties have been reported with the parameterization of such systems and with their maintenance over time \cite{pitts2001optical}, \cite{pitts2002optical}.

A microphone based on Lorentz force was developed by Olson in 1931 \cite{olson1931}. It was made of a metallic ribbon placed between two magnets. The voice vibrated the ribbon inside which the Lorentz force induced an electrical current either recorded or amplified. A hydrophone inspired from this design has been developed in 1969 by Filipczynski, using a thin wire fused on an insulator \cite{filipczynski1969}. Its design has then been improved by Sharf et al. in 1998 by transmitting the electrical signal through electromagnetic coupling, which resulted in an improved resolution, despite a lower sensitivity \cite{sharf1999}. Grasland-Mongrain et al. recently proposed a wire-hydrophone \cite{grasland2012hydroEMarticle}, where a thin electrical wire is moving freely inside the magnetic field of a permanent magnet. The hydrodynamic equations giving the relationship between the movement of the fluid, due to the pressure variations, and the movement of the wire were also proposed under the assumption that the wire diameter remains small compared to the ultrasound wavelength. High spatial resolution was obtained using a tomographic approach through rotations and translations of the hydrophone.

The present article introduces a new Lorentz-force hydrophone design and describes its characteristics. This characterization follows the International Electrotechnical Commission (IEC) 62127-3 standard recommendations \cite{IECstandard2005} by providing the so called basic information (design, size...), the sensitivity, the dynamic range, the linearity and sensitivity to electromagnetic interference, the frequency response, the directionality, and the electric output characteristics. The hydrodynamic model explaining the measurement process is also refined.

\section{Theory}
\begin{figure}[ht]
	\begin{center}
   \includegraphics[width=.8\linewidth]{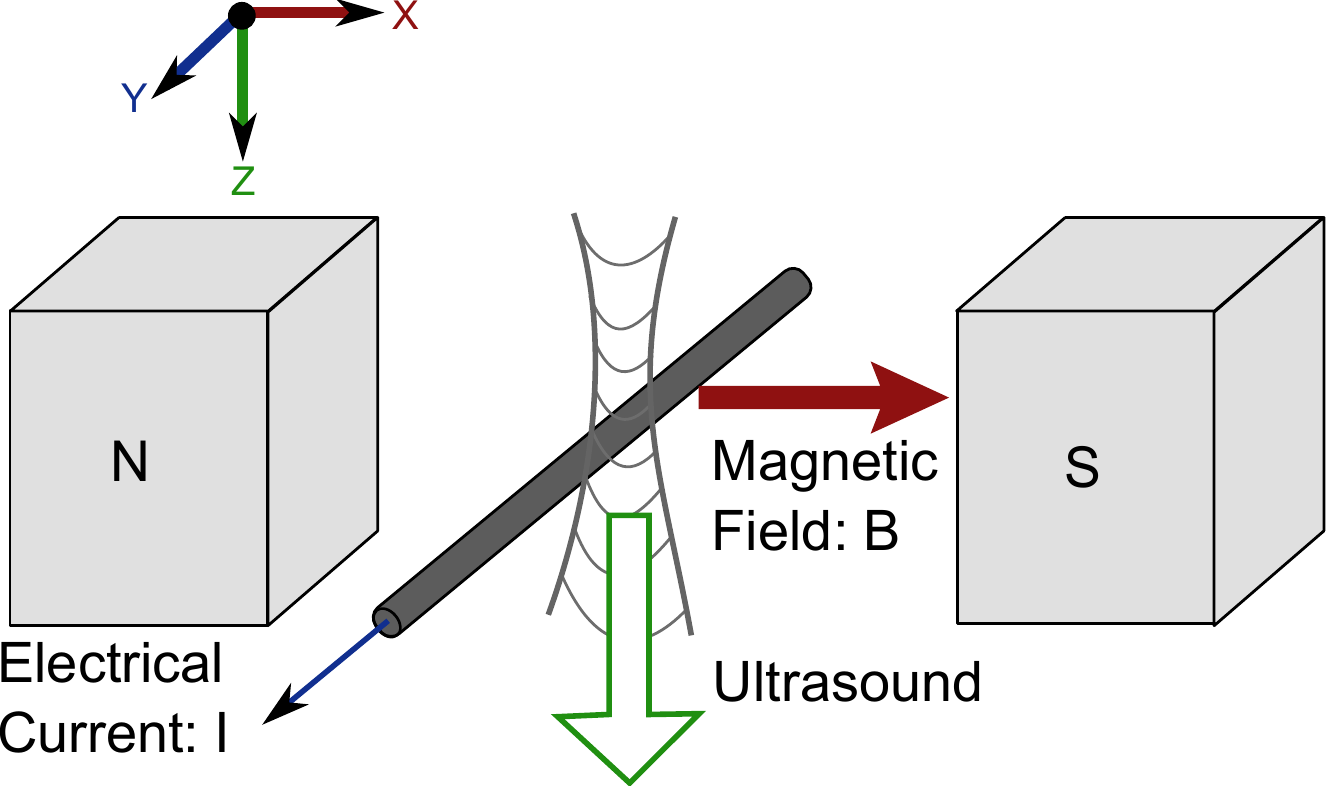}
   \caption{\label{LorentzForce} Principle of the Lorentz-force hydrophone: an ultrasound wave makes an electrical wire move inside a magnetic field, inducing by Lorentz force an electrical current proportional to the wave amplitude.}
	\end{center}
\end{figure}

The principle of the Lorentz-force hydrophone is illustrated in figure (\ref{LorentzForce}); it is made of a thin electrical non-magnetic wire placed in the field of a compound permanent magnet. For clarity, the X axis is aligned along the magnetic field, the Y axis is aligned along the electric wire and the direction of the Z axis follows from the right-hand rule. The magnetic field can be created by a permanent or electrical magnet, but it is shown below that the homogeneity of the field determines how easy it is to exploit this arrangement. The magnetic field is created through a special arrangement of permanent magnets called a Halbach array \cite{halbach1980} which is capable of creating a magnetic field of nearly constant intensity along the whole section of the wire exposed to ultrasound.

The Reynolds number $Re$, defined as $(u D)/\nu$ where $u$ is the fluid velocity, $D$ the cylinder diameter and $\nu$ the kinematic viscosity of the medium, gives the ratio of inertial to viscous forces in a fluid. If this number is sufficiently high and the wire is small compared to the ultrasound wavelength, the velocity $v$ of a cylinder modelling the wire placed in a moving fluid is proportional to the fluid velocity $u$ \cite{landau1987} with a proportionality factor $K$ defined as
\begin{equation}
	\label{theory1}
		K = \frac{2\rho}{\rho_0+\rho},
\end{equation}
where $\rho$ is the fluid density and $\rho_0$ the wire density.

The movement of the conductive wire in the magnetic field induces by Lorentz force a voltage $e$ expressed as
\begin{equation}
	\label{theory2}
		e=\int{(\mathbf{v}(l) \times \mathbf{B}).\mathbf{dl}=\int{\frac{2\rho}{\rho_0+\rho}{u(l)}B dl}}
\end{equation}
where $\mathbf{B}$ is the intensity of the magnetic field and $\mathbf{dl}$ the length element integrated over the wire. If the magnetic field changes too much with location in space, inversing the equation to arrive at the fluid velocity is complicated. It is thus important to obtain a magnetic field as homogeneous as possible along the wire, and it will be assumed here to be constant in the gap of the magnet.

The acoustic velocity field can be retrieved using a tomographic approach \cite{kak1999principles}: measurements performed for different positions and angles of the wire allow the velocity of the fluid $u(x,y)$ to be reconstructed through an inverse Radon transform, which is here expressed as
\begin{equation}
	\label{theory3}
		u(x,y)=\frac{1}{B} \int_0^\pi{E(k,\theta)|k| e^{j2\pi \delta k} d\theta dk}
\end{equation}
where $E$ the Fourier transform of $e$ at location $k,\theta$, $k$ the position and $\theta$ the angle in the imaging plane. 

The method allows the Z-component of the local velocity vector to be extracted depending on the directionality of the Lorentz-force. 
	
If the ultrasound wave can be locally considered as a plane wave, the fluid velocity $u$ and pressure $p$ are related by the equation $u=p/(\rho c)$, where $c$ is the ultrasound wave speed. The pressure can then be calculated as
\begin{equation}
\label{theory4}
		p(x,y)=\frac{\rho c}{B} \int_0^\pi{E(k,\theta)|k| e^{j2\pi \delta k} d\theta dk}
\end{equation}

In practice, the hydrophone would be placed in the imaging plane so as to measure the induced signal at different positions and angles (depending on the desired spatial resolution and quality of reconstruction). An inverse Radon transform performed directly on the voltage data acquired would produce a map of the velocity field (or pressure field via the plane wave equation).

The hydrodynamic model can be refined by looking at the directional response of the hydrophone. The induced voltage is different if the axis of rotation for the directional response is along the X or Y direction.

\begin{figure}[ht]
   \begin{minipage}[c]{.45\linewidth}
		\begin{center}
		(a)
	   		\includegraphics[width=.9\linewidth]{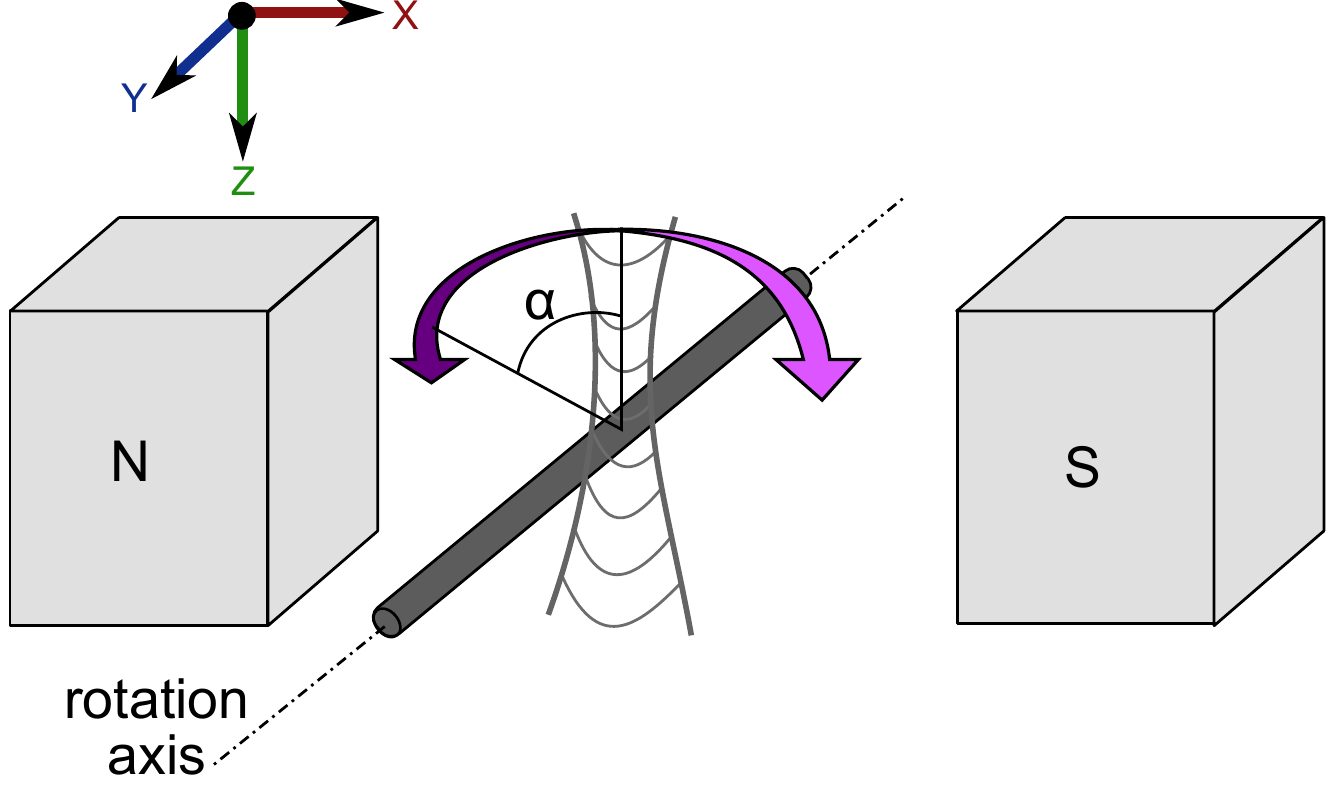}
		\end{center}
   \end{minipage} \hfill
   \begin{minipage}[c]{.45\linewidth}
		\begin{center}
		(b)
	   		\includegraphics[width=.9\linewidth]{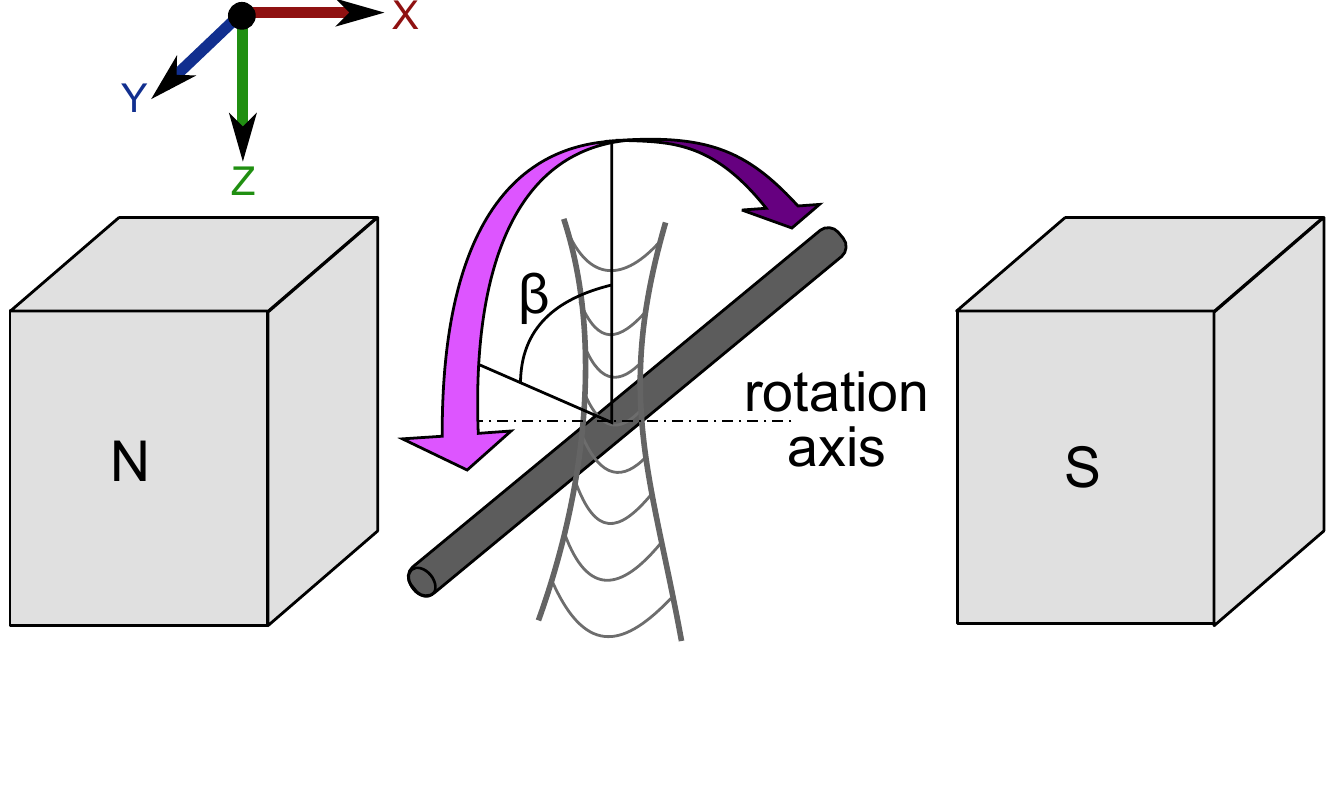}
		\end{center}
   \end{minipage}
	   		\caption{\label{Direction_schema} (a) Directional response for a rotation around the Y axis, with $\alpha$ the angle between the ultrasound propagation direction and the Z axis; (b) Directional response for a rotation around the X axis, with $\beta$ the angle between the ultrasound propagation direction and the Z axis.}
\end{figure}

For rotation around the Y axis as illustrated in figure \ref{Direction_schema}-(a), the expression for the voltage $e$ induced by the Lorentz force indicates a sinusoidal dependency, expressed as 
\begin{equation}
	\label{direction0}
		e = \int{(\mathbf{v} \times \mathbf{B}).\mathbf{dl}} = \int{vB\cos(\alpha)dl}
\end{equation}
where $\alpha$ is the angle formed with the Z axis in the XZ plane.

The directional response with a rotation around the X axis as illustrated in figure \ref{Direction_schema}-(b) is also different. Due to the wire geometry, the current can be induced only in the Y axis direction, and is solely generated by the Z-component of the movement of the wire. The voltage $e$ can then be expressed as
\begin{equation}
	\label{direction1}
		e = \int{(\mathbf{v} \times \mathbf{B}).\mathbf{dl}} =  \int{v \cos(\beta) B dl}
\end{equation}
where $\beta$ is the angle formed with the Z axis in the XY plane.

In this second case, the wire can be exposed simultaneously to nodes and antinodes of pressure depending on the ultrasound waveform and the induced voltage is averaged. Assuming that the ultrasound waveform is sinusoidal, the fluid velocity $u$ is equal to
\begin{equation}
	\label{direction2}
	 u(\xi) = u_0 \cos( \frac{2\pi}{\lambda}\xi )
\end{equation}
where $u_0$ is the maximum velocity of the fluid and $\xi$ = $d \tan(\beta)$, $d$ being the ultrasound beam width. Combining equations (\ref{direction1}) and (\ref{direction2}) and calculating the integral between $0$ and $L$ leads for the following expression of the induced voltage
\begin{equation}
	\label{direction3}
	 e = \frac{u_0 B}{d} \sinc(\frac{2 \pi d}{\lambda} \tan(\beta))
\end{equation}

However this equation is a particular case matching the shape of the emitted ultrasound, contrary to the other directional response.

\section{Material and Methods}

\subsection{Hydrophone design}
\begin{figure}[ht]
   \begin{minipage}[c]{.45\linewidth}
		\begin{center}
		(a)
	   		\includegraphics[width=.9\linewidth]{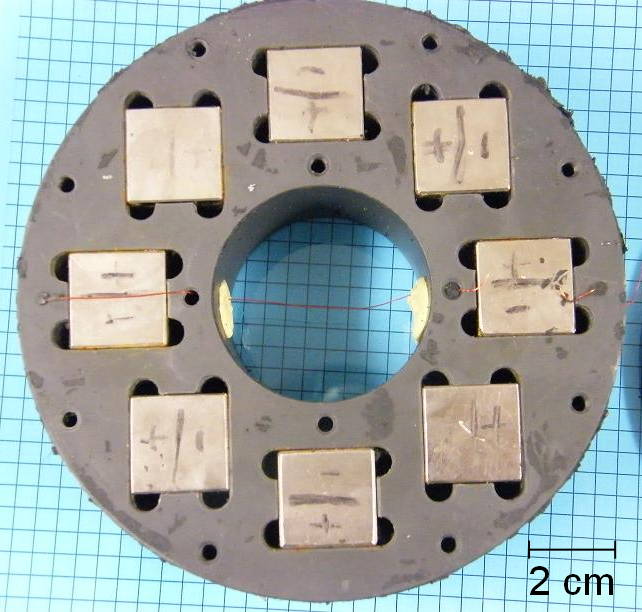}
		\end{center}
   \end{minipage} \hfill
   \begin{minipage}[c]{.45\linewidth}
		\begin{center}
		(b)
	   		\includegraphics[width=.9\linewidth]{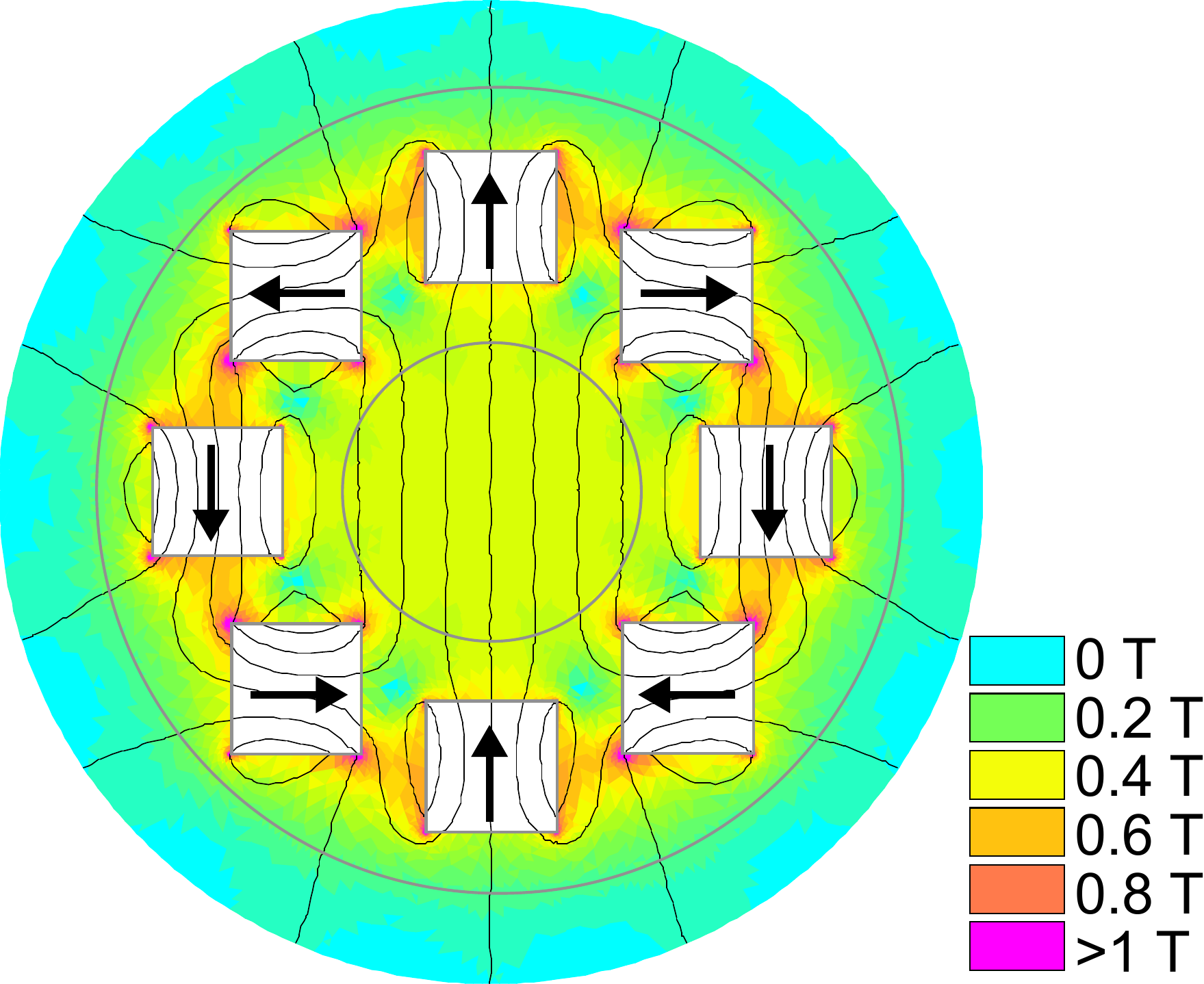}
		\end{center}
   \end{minipage}
	   		\caption{\label{HydrophoneEM} (a) Picture of the Lorentz-force hydrophone, with magnets placed as an Halbach array. Note that the wire was taut with two screws placed on the PVC housing before use ; (b) Magnetic field simulation with flux line, giving a quasi homogenous value of 0.17 $\pm$ 0.04 T inside the cylinder.}
\end{figure}

A Lorentz-force hydrophone was created from eight 35x20x20 mm$^3$ neodymium magnets (HKCM Engineering, Eckernfoerde, Germany) placed regularly in a cylindrical PVC housing with 12 cm (external) and 4.6 cm (internal) diameters, as shown in figure (\ref{HydrophoneEM})-(a). The resulting Halbach array \cite{halbach1980} creates a homogeneous magnetic field inside the cylinder. The magnetic field was simulated in two dimensions using a finite-elements software (Finite Element Method Magnetics 4.2, {http://www.femm.info}), which is illustrated in figure (\ref{HydrophoneEM})-(b), and measured using a manual teslameter (Model 612003, Unilabs, Clichy-la-Garenne, France).

\begin{figure}[ht]
		\begin{center}
	   		\includegraphics[width=.9\linewidth]{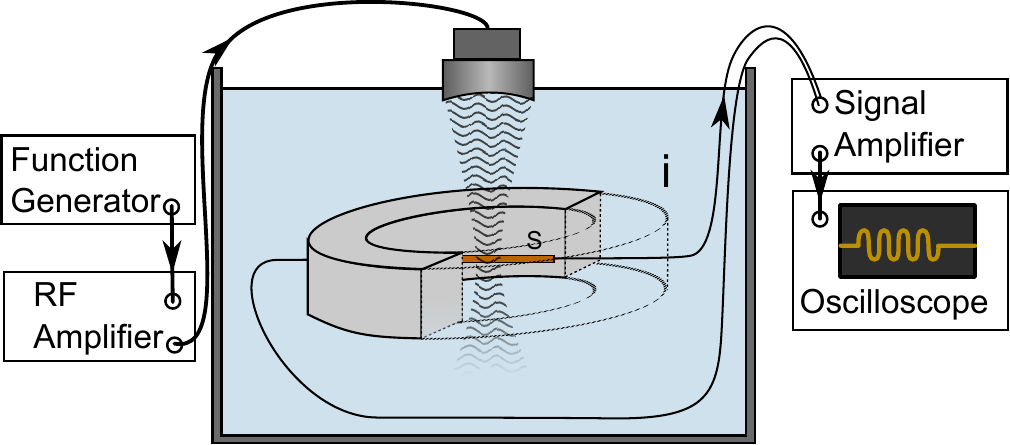}
	      \caption{\label{HydroEM_exp} Experiment with the Lorentz-force hydrophone. A function generator with an RF amplifier drives a focused transducer with the Lorentz force placed at the focal point. The induced signal is amplified and observed with an oscilloscope.}
		\end{center} 
\end{figure}

The sensitive part of the hydrophone was essentially consisted of an insulated copper wire placed in the middle of the cylinder. The wire had a diameter of 200 \micro m and was kept straight by applying a small tension, the magnitude of which is later shown to have little effect. The impedance between the two extremities was 0.8$\pm$0.2 $\Omega$ at 0.5 MHz. The hydrophone was connected to a 1 MV.A$^{-1}$ current amplifier (HCA-2M-1M, Laser Components, Olching, Germany) with an input impedance of 250 $\Omega$ at 0.5 MHz. The signal was then observed and measured using an oscilloscope (WaveSurfer 422, LeCroy, Chestnut Ridge, NY, USA). A typical experiment is illustrated in figure (\ref{HydroEM_exp}).

The function generator was a Tektronix AFG 3022B (Tektronix, Beaverton, OR, USA) which was connected to a 200W power amplifier (200W LA200H, Kalmus Engineering, Rock Hill, SC, USA). The ultrasound transducer had a custom-made housing with an air-backed piezoelectric membrane (piezoelectric ceramic PZ-28, Meggitt, Kvistgraad, Denmark) at a central frequency of 0.5 MHz. Its piezoelectric element was spherical with a diameter of 50 mm and a radius of curvature of 50 mm. The beam width at 0.5 MHz was 8 mm at the focus.

For frequency and pressure linearity experiments, characteristics were compared with the measurements of a calibrated lipstick hydrophone with a 20 dB preamplifier (HGL-200-1113 with AH2010 preamplifier, Onda Corporation, Sunnyvale, CA, USA).

\subsection{Wire tension}
The hydrodynamic model assumes that the wire follows the movement of the surrounding fluid perfectly, but this assumption could be invalid if the wire tension were too high, preventing large oscillations. To verify this, one section of the study aims to evaluate the influence of the wire tension on the induced signal.

The function generator was set to create a 1 V peak-to-peak, 0.5 MHz signal with 4 sinusoids per pulse at a transmit repetition frequency of 250 Hz. These pulses were amplified by the power amplifier before being sent to the transducer. This generated a pressure of 3 MPa at the focal point where the wire in the Lorentz-force hydrophone was placed, perpendicularly to the ultrasound beam. 

As the hydrophone was lying horizontally, the wire had a catenary shape due to its own weight. This shape was described three hundred years ago by Bernoulli, Huyghens and Leibniz \cite{routh1891}. The mechanical tension in the wire was calculated from the catenary equation
\begin{equation}
	y(x) = a \cosh(x/a)+c
	\label{tension}
\end{equation}
where $y$ is the height of the wire at the $x$ coordinate, $a = \frac{H}{\lambda_0 g}$ (where $H$ is the horizontal component of the tension, $\lambda_0$ the mass per unit length and $g$ the acceleration due to gravity) and $c$ height of the wire extremities.

The ultrasound beam was focused at the lowest point of the wire and the induced signal in the hydrophone was recorded. Then, as the mechanical tension was increased by rotating the two screws maintaining it, the wire became straighter and the lowest point moved closer to the transducer. The delay between ultrasound emission by the transducer and its reception by the hydrophone changed and was consequently measured. Assuming a speed of sound of 1470 m.s$^{-1}$, the corresponding displacement was computed as the wire tension was increased. The tension was then divided by the cross-sectional area (a 70 \micro m diameter disc) to obtain the stress in pascals, making it simpler to extend the results to other types of wire. The tension was increased gradually until disruption, estimated at 220 MPa from the literature \cite{wildi2000}.

To quantify the effect of tension on the signal, we calculated the correlation coefficient defined as
\begin{equation}
	\label{coefCorrFormula}
	Q = \frac{  \sum{\Psi _1 \Psi _2}  }   {\sqrt{   \sum{\Psi _1 ^2}\sum{\Psi _2 ^2}   }},
\end{equation}
where $\Psi_1$ and $\Psi_2$ are two signals. This coefficient is equal to 1 when $\Psi_1$ and $\Psi_2$ are identical, -1 when exactly opposite and 0 when uncorrelated. In the experiment addressing wire tension, the correlation coefficient was calculated between the signals measured at different stresses, using the signal with the lowest tension as a reference. Only the signal between 50 and 100 \micro s (plus a few microseconds of delay depending on the wire displacement), which is approximately the incidence time of the wave on the wire, was used for computation.

\subsection{Frequency response}
Unlike source transducers, most hydrophones have no electrical impedance resonance peaks in their operability bandwidth. To check for the absence of such peaks, the real and imaginary impedances versus frequency were measured.

The Lorentz-force hydrophone was immersed in water and directly connected, without using any current amplifier, to an impedance meter (ZVB-4 Vector Network Analyzer, Rohde Schwarz GmbH, Munich, Germany) with 50 $\Omega$ input impedance.

Another experiment was then performed to investigate the hydrophone acoustic frequency response.

The function generator was set to generate a 1 V peak-to-peak signal with 4 sinusoids per pulse at a repetition frequency of 100 Hz, amplified by the power amplifier, applied to a 2 MHz wide-band spherical transducer with a diameter of 50 mm and a radius of curvature of 210 mm (Imasonic SAS, Voray sur l'Ognon, France). The pulse frequency was increased from 100 kHz to 5 MHz in 20 kHz steps. This generated a pressure of approximately 3 MPa at the focal point where the wire in the Lorentz-force hydrophone was placed, perpendicularly to the ultrasound beam.

Four different wire diameters were tested: 70$\pm$7 \micro m, 107$\pm$7 \micro m, 217$\pm$7 \micro m and 400$\pm$7 \micro m. Due to its low cutoff frequency (-3 dB at 2 MHz), the HCA-2M-1M amplifier was replaced by a wide-band current amplifier (HCA-10M-100k, Laser Components, Olching, Germany).

\subsection{Directional Response}
The directional model presented in the theory section was tested with an experiment where the ultrasound beam was not orthogonal to the magnetic field or the wire hydrophone direction.

The generator emitted a 1 V peak-to-peak signal with 4 sinusoids per pulse at a pulse repetition frequency of 250 Hz, amplified by the power amplifier, to the transducer. This generated a pressure of 3 MPa at the focal point where the wire of the Lorentz-force hydrophone was placed, perpendicularly to the ultrasound beam.

The angle between the transducer and the hydrophone ranged from -35$^o$ to +35$^o$ with steps of 0.5$^o$ in two directions, with the rotations performed around the X and Y axis. Larger angles could not be evaluated due to the current hydrophone design. Eight sets of readings were taken in each direction.

\subsection{Sensitivity, dynamic range, linearity and electromagnetic interference}
One of the advantages of the Lorentz-force hydrophone is its robustness at high pressure. The following experiment aimed to evaluate the range of pressure measurable by the hydrophone and the linearity of the response, and observe the impact on the hydrophone as the amplitude is increased.

The generator emitted a 0.5 MHz signal with 4 sinusoids per pulse at a pulse repetition frequency of 100 Hz. This signal was amplified and delivered to the transducer. The wire of the Lorentz-force hydrophone was placed at the focal point, perpendicularly to the ultrasound beam centre line.

Three different amplification conditions were tested: with no amplifier, with the 200W power amplifier, and with a 500 W power amplifier (1040L, Electronics and Innovation Engineering, Rochester, NY, USA), to obtain a dynamic range from 10 kPa to 10 MPa. At high pressure, the output of the HCA-2M-1M signal amplifier saturated at 4 V, so the experiment was repeated without the current amplifier, and the Lorentz-force hydrophone was directly connected to the oscilloscope with an 1 M$\Omega$ impedance. As the measured signal derives from the physical integration of the Lorentz effect over the whole exposed section of the wire, the measure was normalized by the ultrasound beam width.

At high ultrasound pressure, the main potential risks are heating and cavitation induced by sonication. However if the wire diameter is small compared to the ultrasound wavelength, a very small part of the ultrasound energy will be absorbed by the wire and the resulting increase in temperature will be low. Moreover, the wire is constantly cooled by the surrounding medium, further reducing the temperature rise. A global increase in its temperature would change the wire resistance which would nevertheless remain small compared to the input impedance of the amplifier. Regarding cavitation, the mechanical stresses induced by a cavitation cloud on the wire could damage it and alter the measured signals.

An experiment was thus performed using specific equipment described below to obtain a stable cavitation cloud over a long period of time. The aim of this experiment was to compare the signal before and after being submitted to strong-cavitation clouds. A computer equipped with a transmission/reception module (NI5781R module, National Instruments, Austin, TX, USA) produced an excitation signal, amplified by a 69 dB power amplifier (Prana RD, Brive, France), applied to a 550 kHz transducer with a 10 cm diameter and focal distance of 10 cm. To evaluate the intensity of the cavitation, the cavitation index was computed, defined as $CI=20\log(\left|E\right|)$ where $E$ is the Fourier transform of the signal between .1 and 3 MHz \cite{desjouy2013}. The measurement was then made in three steps:
\begin{enumerate}
	\item Excitation with a signal of 0.05 V, 28 pulses per burst, with a pulse repetition frequency of 4 Hz to the transducer. The pressure was recorded by the Lorentz-force hydrophone placed at the focal point.
	\item Excitation with a signal of 0.1\% duty cycle, 4 Hz pulse repetition frequency, 2 V mean amplitude controlled by the computer module through a feedback loop, and resulting in a focal peak-to-peak pressure of around 15$\pm$1 MPa. The cavitation index was equal to 15 dB. Cavitation clouds as large as 1 cm in diameter were observed on the hydrophone wire. As cavitation is a random process, the Lorentz force did not registered the signal.
	\item Same excitation as (a) with a signal of 0.05 V, 28 pulses per burst, with a pulse repetition frequency of 4 Hz to the transducer. The pressure was recorded by the Lorentz-force hydrophone placed at the focal point.
\end{enumerate}
The change of the signal due to the cavitation was calculated using the correlation coefficient $Q$ as defined in equation \ref{coefCorrFormula} between the signal before and the signal after cavitation.

\section{Results}
The simulation has given a magnetic field strength inside the cylinder equal to $0.17 \pm 0.04$ T. This value was confirmed by the measure performed with the manual teslameter.

\subsection{Wire tension}
\begin{figure}[ht]
		\begin{center}
	   \includegraphics[width=.90\linewidth]{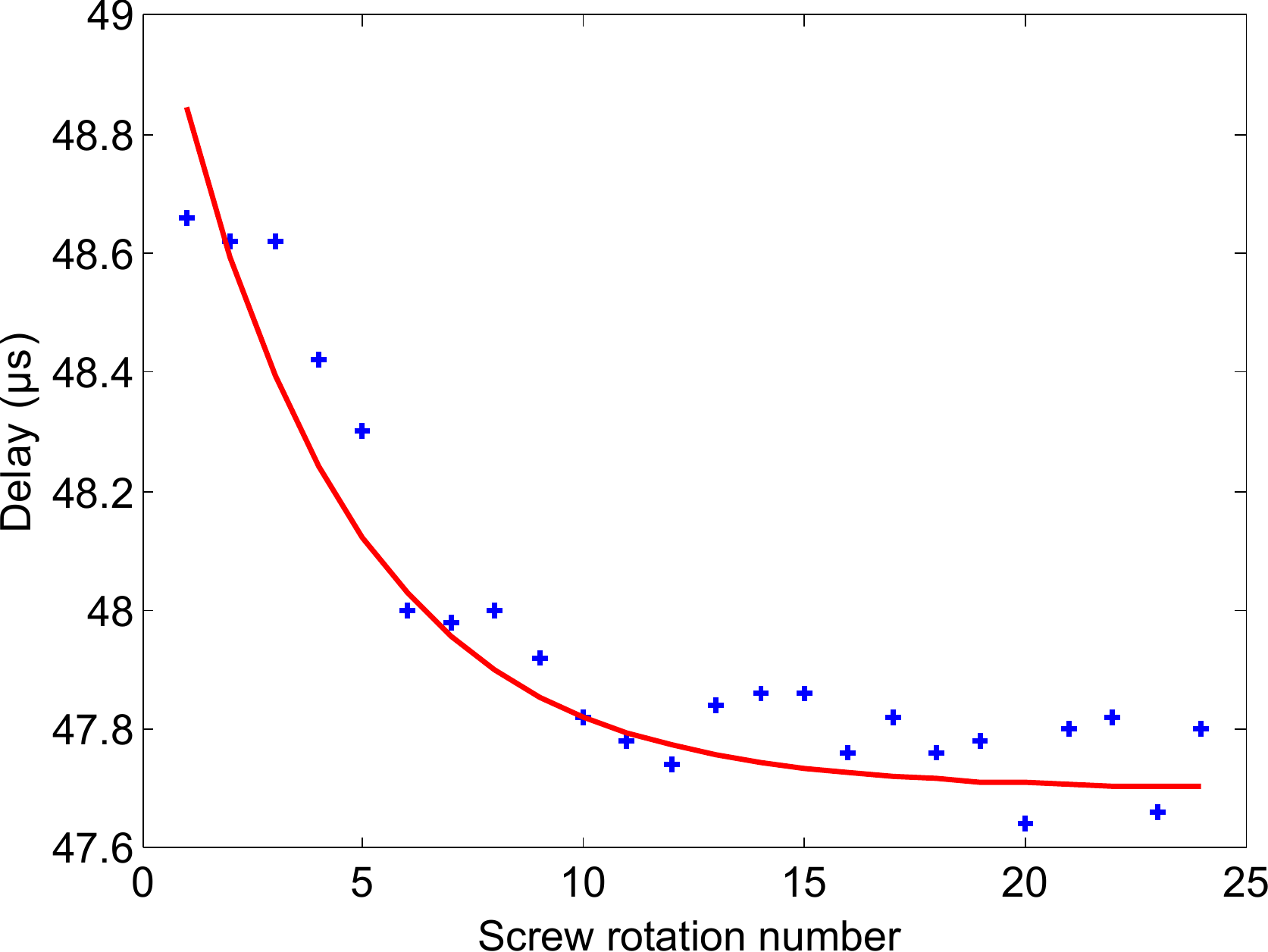}
	   \caption{\label{ResultatTension} As the screw maintaining the wire are turned, the wire becomes tauter and the delay between the ultrasound emission and their reception by the hydrophone was decreasing. Delay for each screw rotation (blue crosses); Exponential fit of the data (red line).}
		\end{center} 
\end{figure}

The delay between the ultrasound emission and its reception by the hydrophone at each increase of tension is illustrated in figure (\ref{ResultatTension}). The delay is decreasing with increasing tension, but the ten last changes were almost undetectable: they were guessed by extrapolation from the first measurements and from the rupture stress test with an inverse exponential fit on the data. The maximal displacement was found to be equal to 1.5 mm (delay of 1 \micro s with a wave velocity of 1470 m.s$^{-1}$), indicating that the initial stress was equal to 300 kPa. 

The table \ref{coefCorrTension} shows the estimated wire stress versus the correlation coefficient as defined in (\ref{coefCorrFormula}). The correlation coefficient is always above 0.97, showing a high correlation between the successive measures.

\begin{table}
	\caption{Correlation coefficient between signals with increasing wire stress.}
	\begin{tabular}{|c|c|}
		\hline
		Estimated Wire Stress (MPa) & Correlation coefficient Q \footnote{Reference signal is for wire stress of 0.3 MPa.} \footnote{Correlation coefficient is always above 0.97, meaning, that high correlation is found between the signals at different tensions.}\\
		\hline
		0.3 &1.0000\\
		0.4 &0.9843\\
		0.7 &0.9840\\
		1   &0.9751\\
		2   &0.9765\\
		3   &0.9768\\
		5   &0.9657\\
		7   &0.9760\\
		10  &0.9802\\
		20  &0.9786\\
		40  &0.9775\\
		50  &0.9711\\
		80  &0.9874\\
		100 &0.9833\\
		200 &0.9829\\
		\hline
	\end{tabular}
	\label{coefCorrTension}
\end{table} 

\subsection{Frequency response}
\begin{figure}[ht]
		\begin{center}
	   \includegraphics[width=.90\linewidth]{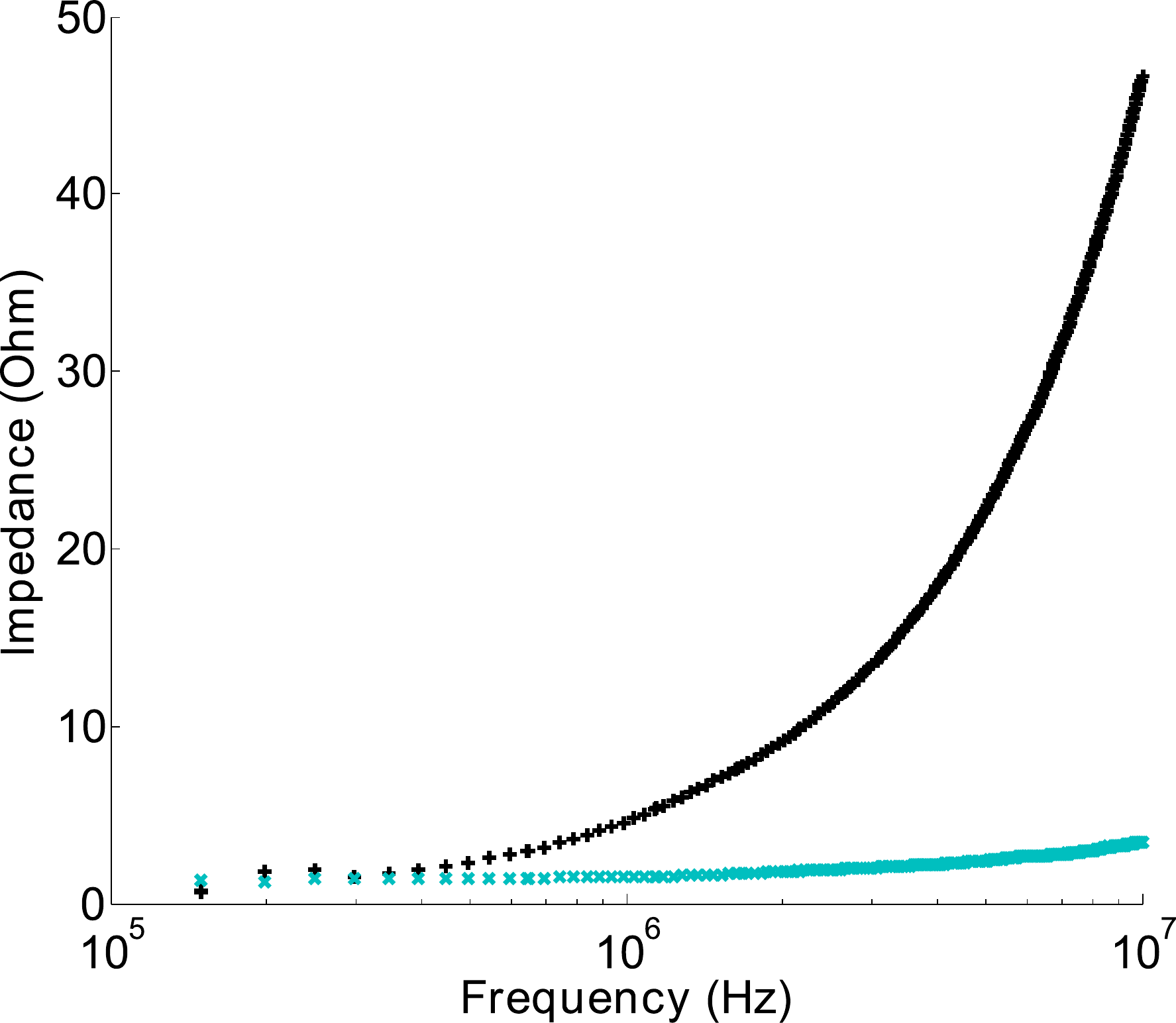}
	   \caption{\label{Impedance10} End-of-cable real and imaginary impedances of the Lorentz-force hydrophone versus frequency from 0.15 to 10 MHz. The real impedance increases from 0.71 to 4.3 $\Omega$ and the imaginary part from 0.89 to 63 $\Omega$. No peak of resonance can be seen.}
		\end{center}
\end{figure}

End-of-cable real and imaginary impedances versus frequency are plotted in figure (\ref{Impedance10}), from 0.15 to 10 MHz. The real impedance increases from 0.71 to 4.3 $\Omega$ and the imaginary part from 0.89 to 63 $\Omega$. The lowest peak of resonance is observed at 30 MHz.

\begin{figure}[ht]
		\begin{center}
   		\includegraphics[width=.90\linewidth]{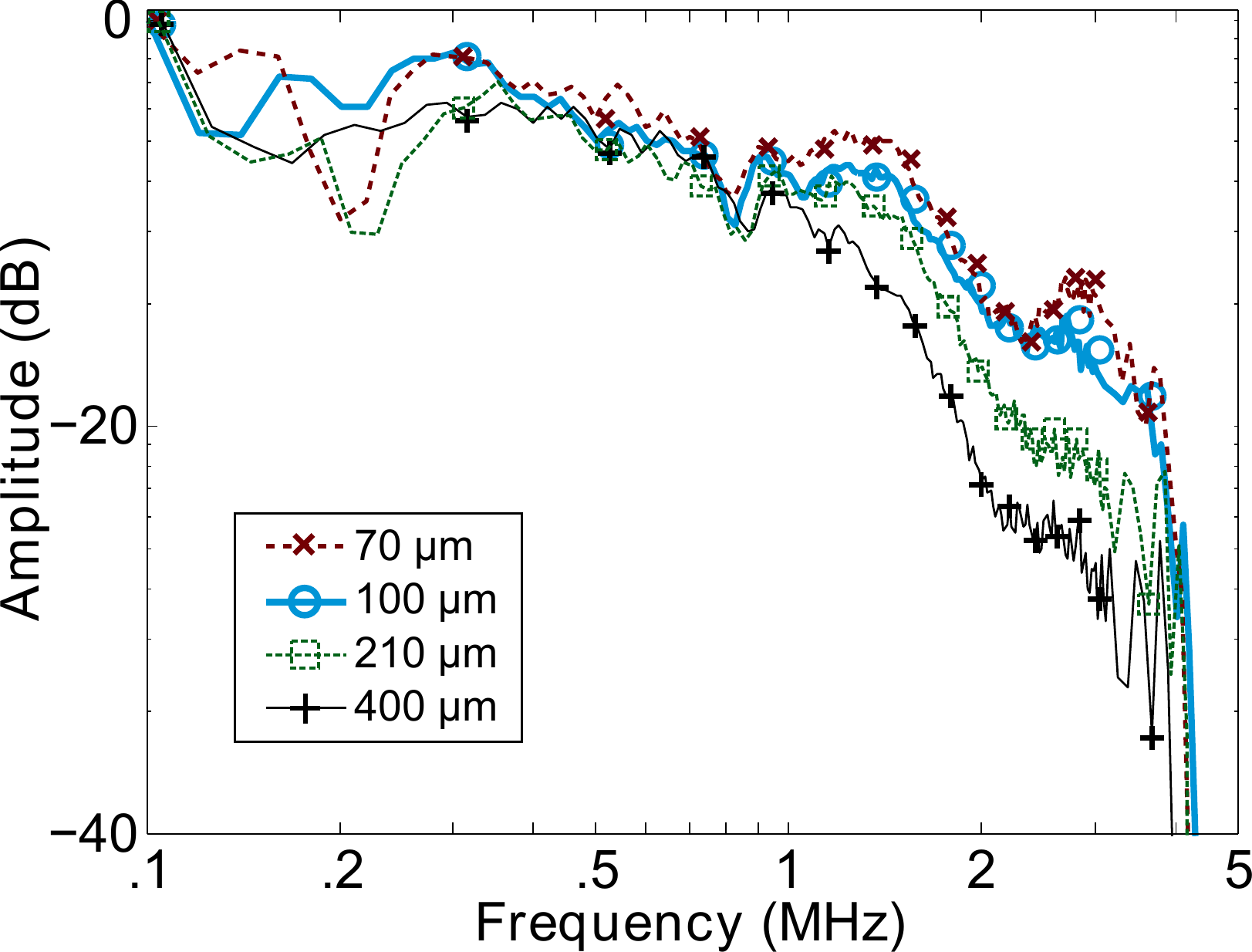}
   		\caption{\label{ReponseFrequentielle} Frequency response of the Lorentz-force hydrophone with four different wire diameter: 70, 100, 200 and 400 \micro m. The upper cutoff frequency is lower for thicker wires, from 2 MHz for the 70 \micro m wire to 1 MHz for the 400 \micro m wire. The signal-to-noise ratio above 3 MHz is too low to get any significant value.}
		\end{center}
\end{figure}

The frequency response between 100 kHz and 2 MHz acquired by the Lorentz-force hydrophone with four different wire diameters (70, 100, 200 and 400 \micro m) divided by the signal acquired by the calibrated piezoelectric hydrophone is plotted in figure (\ref{ReponseFrequentielle}). The upper cutoff frequency is lower for thicker wires, from 2 MHz for the 70 \micro m wire to 1 MHz for the 400 \micro m wire. The signal-to-noise ratio above 3 MHz is too low to get any significant value. 

\subsection{Directional Response}
\begin{figure}[ht]
		\begin{center}
   		\includegraphics[width=.90\linewidth]{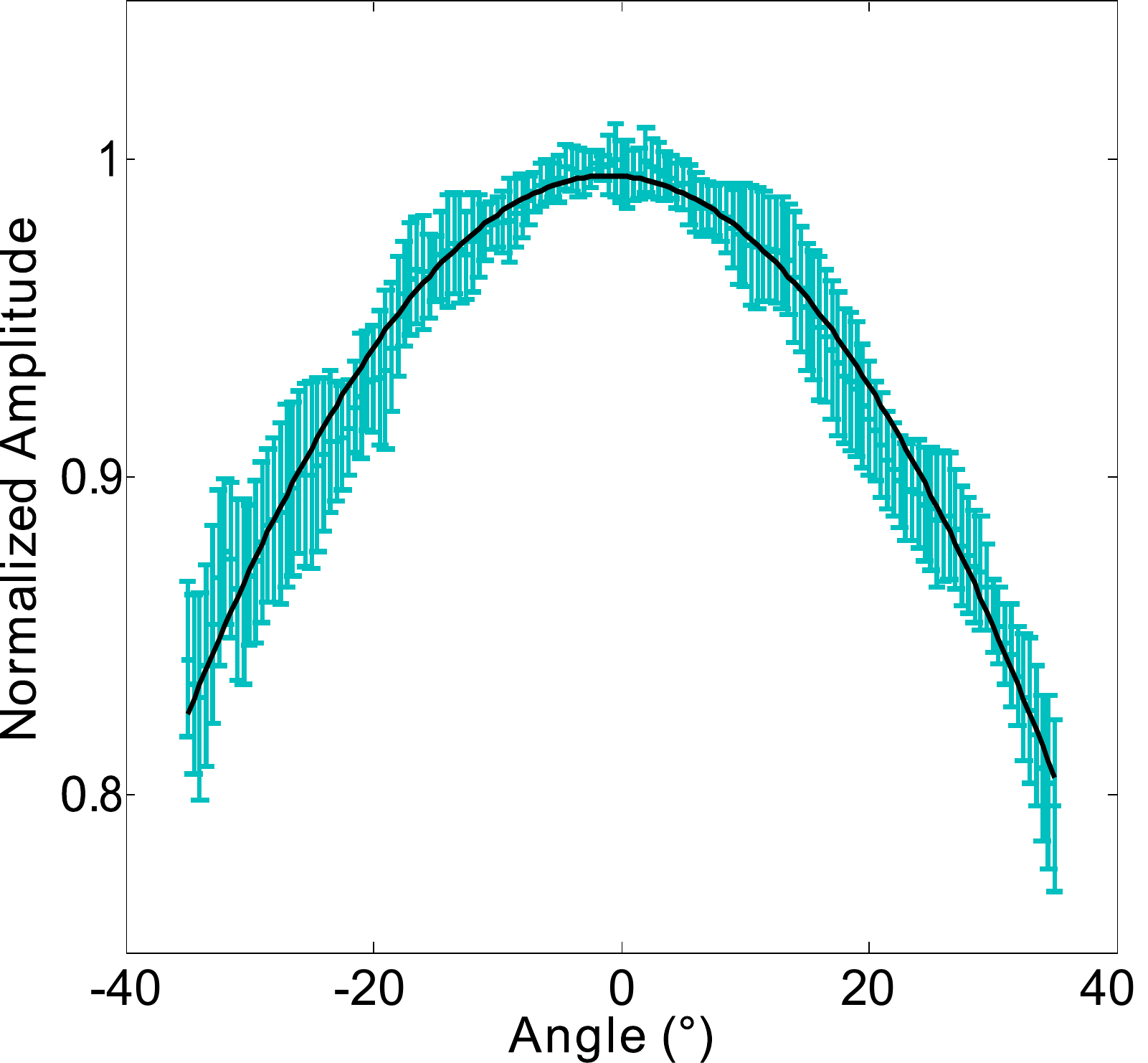}
   		\caption{\label{DirectionVerticale} Normalized amplitude versus angle measurements around the Y direction; Plot of function cos($\alpha$) versus angle $\alpha$. The fit is in the uncertainty range.}
		\end{center} 
\end{figure}

\begin{figure}[ht]
		\begin{center}
   		\includegraphics[width=.90\linewidth]{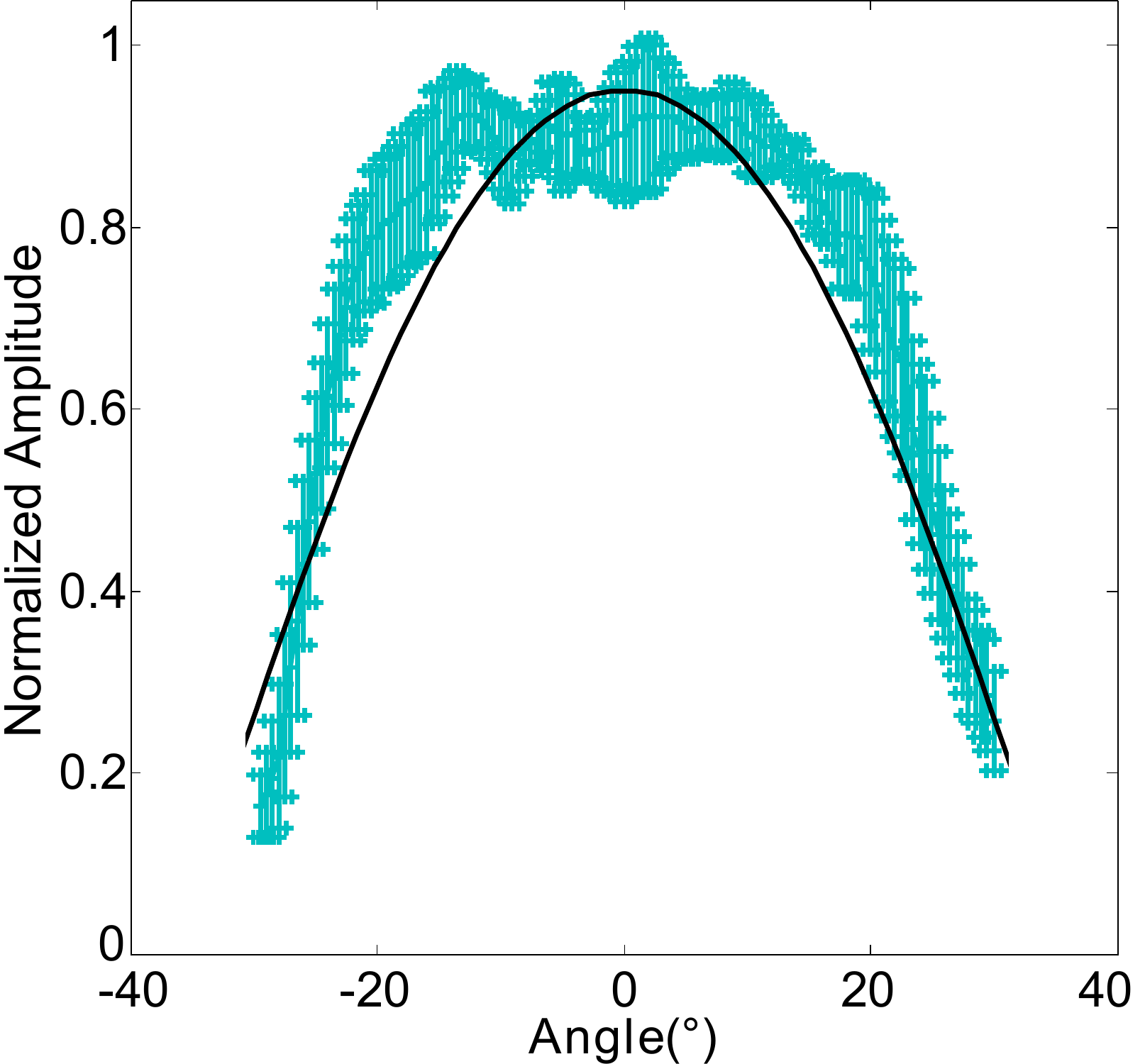}
		  \caption{\label{DirectionHorizontale} Normalized amplitude versus angle measurements around the X direction; Plot of function $sinc(\frac{2\pi}{\lambda}tan(\beta))$ versus angle $\beta$. The fit is mostly in the uncertainty range.}
		\end{center}
\end{figure}

The normalized amplitude of the measured signal versus incidence angle is plotted in figure (\ref{DirectionVerticale}) with axis of rotation in the Y direction and in figure (\ref{DirectionHorizontale}) with axis of rotation in the X direction. For each figure error bars denote the standard deviation obtained from the eight acquisitions. Cosinus of the angle predicted by equation (\ref{direction0}) is plotted in figure (\ref{DirectionVerticale}). Similarly, the amplitude calculated using equation (\ref{direction3}) is plotted in figure (\ref{DirectionHorizontale}).

\subsection{Sensitivity, dynamic range, linearity and electromagnetic interference}
\begin{figure}[ht]
		\begin{center}
	   		\includegraphics[width=.95\linewidth]{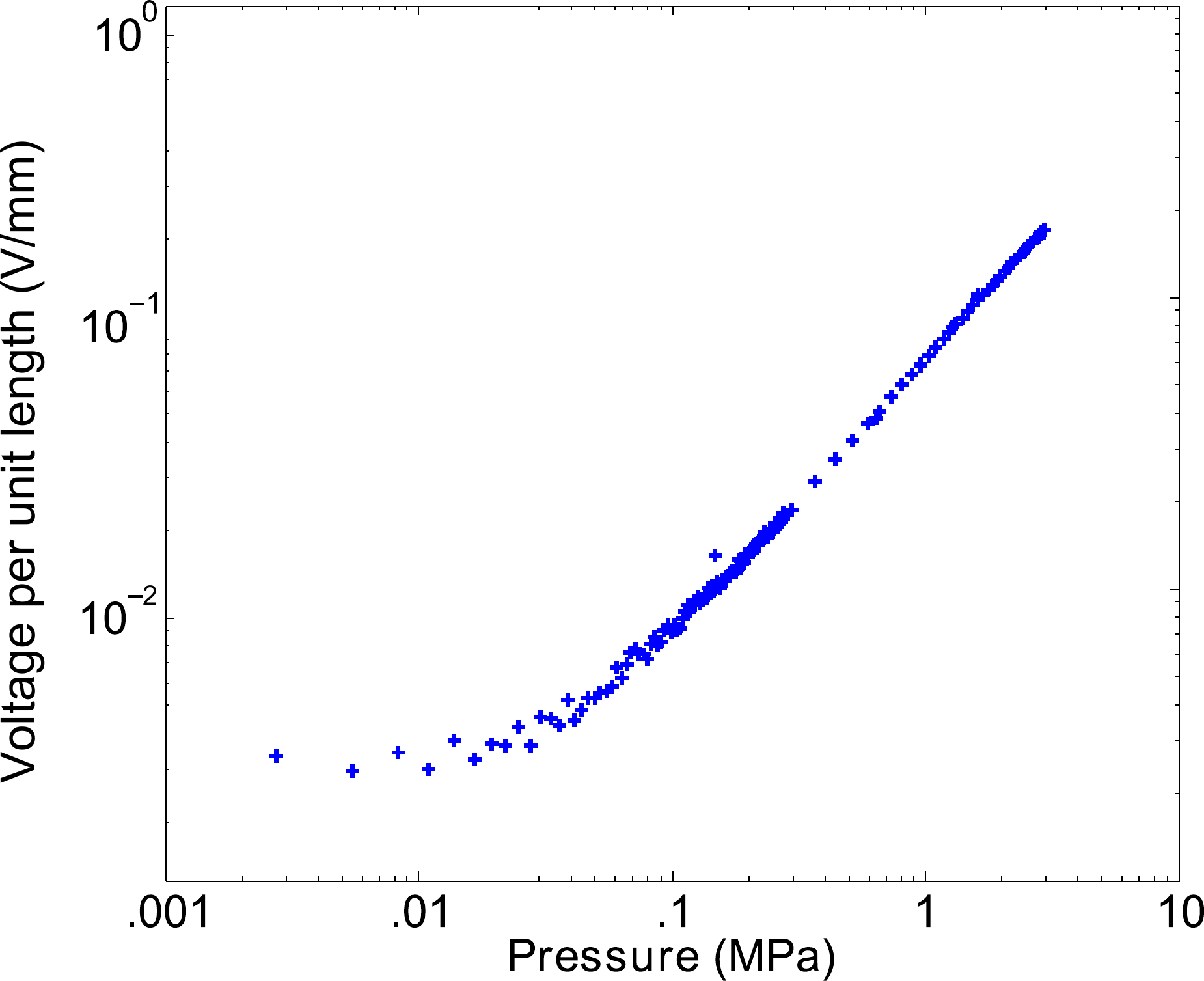}
	   		\caption{\label{PressureWithHCA} End-of-cable voltage per unit length over pressure measured with the signal amplifier. Signal is linear from 50 kPa to 2 MPa.} 
		\end{center}
\end{figure}

\begin{figure}[ht]
		\begin{center}
	   		\includegraphics[width=.95\linewidth]{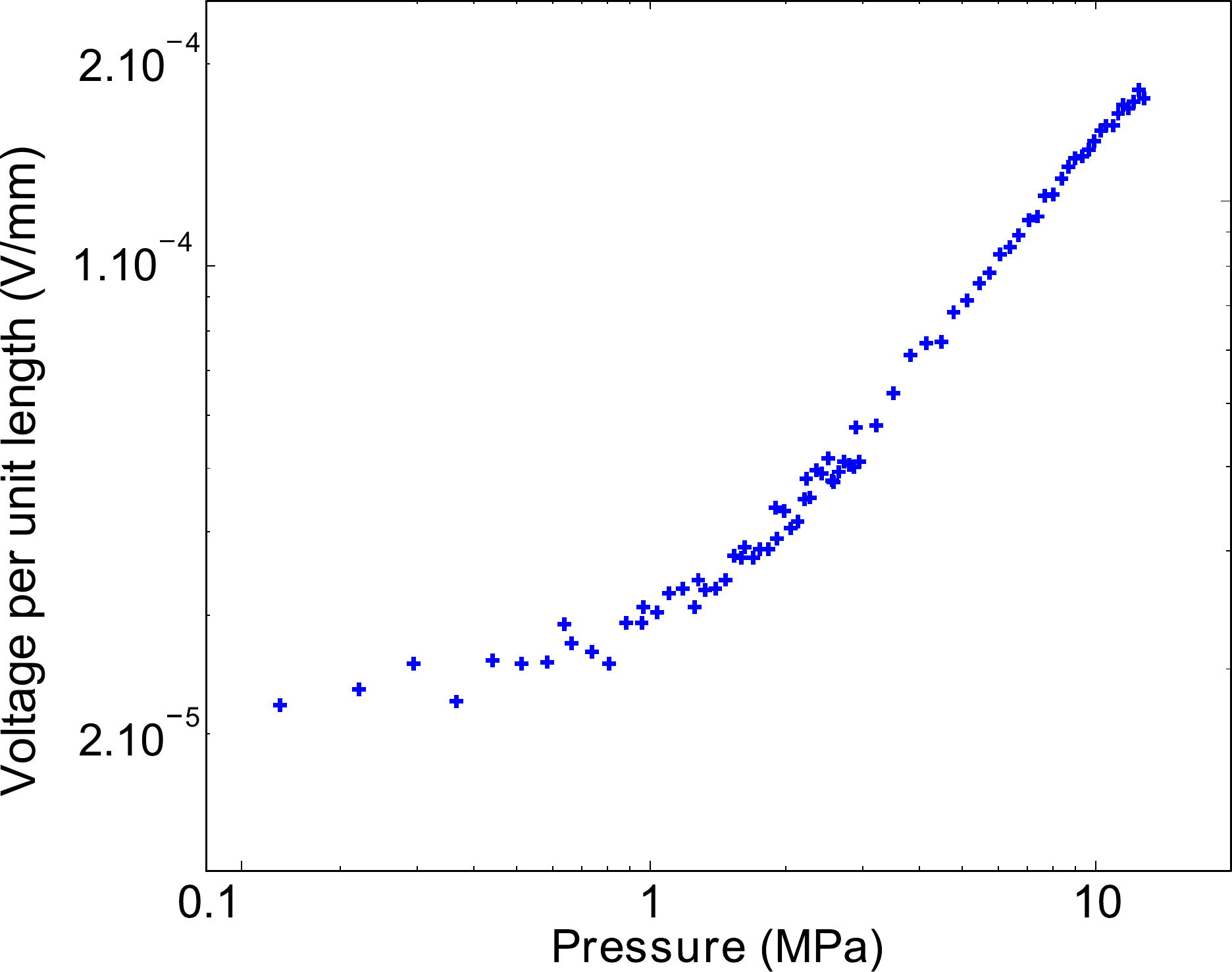}
	   		\caption{\label{PressureWithoutHCA} End-of-cable voltage per unit length over pressure measured without the signal amplifier. Signal is linear from 1 MPa to 10 MPa.}
		\end{center}
\end{figure}

Results obtained with and without the power amplifiers have been plotted in figure (\ref{PressureWithHCA}). The measurements performed without the output current amplifier (due to its saturation at the highest powers) are plotted in figure (\ref{PressureWithoutHCA}). The output signal is linear with ultrasound pressure over a range from 50 kPa to 10 MPa. A linear fitting on the data gives a sensitivity of $1.4 \pm 0.1\ 10^{-11}$ V.mm$^{-1}$.Pa$^{-1}$, or -217 to -216 dB re \micro V.mm$^{-1}$.Pa$^{-1}$, with a determination coefficient $R^2$ equal to 0.997 without the signal amplifier. Similarly, a linear fitting gives a sensitivity of $8.6 \pm 0.1\ 10^{-8}$ V.mm$^{-1}$.Pa$^{-1}$, or -141.4 to -141.1 dB re \micro V.mm$^{-1}$.Pa$^{-1}$, with $R^2$ equal to 0.9981 with the signal amplifier.

\begin{figure}[ht]
	   \includegraphics[width=.90\linewidth]{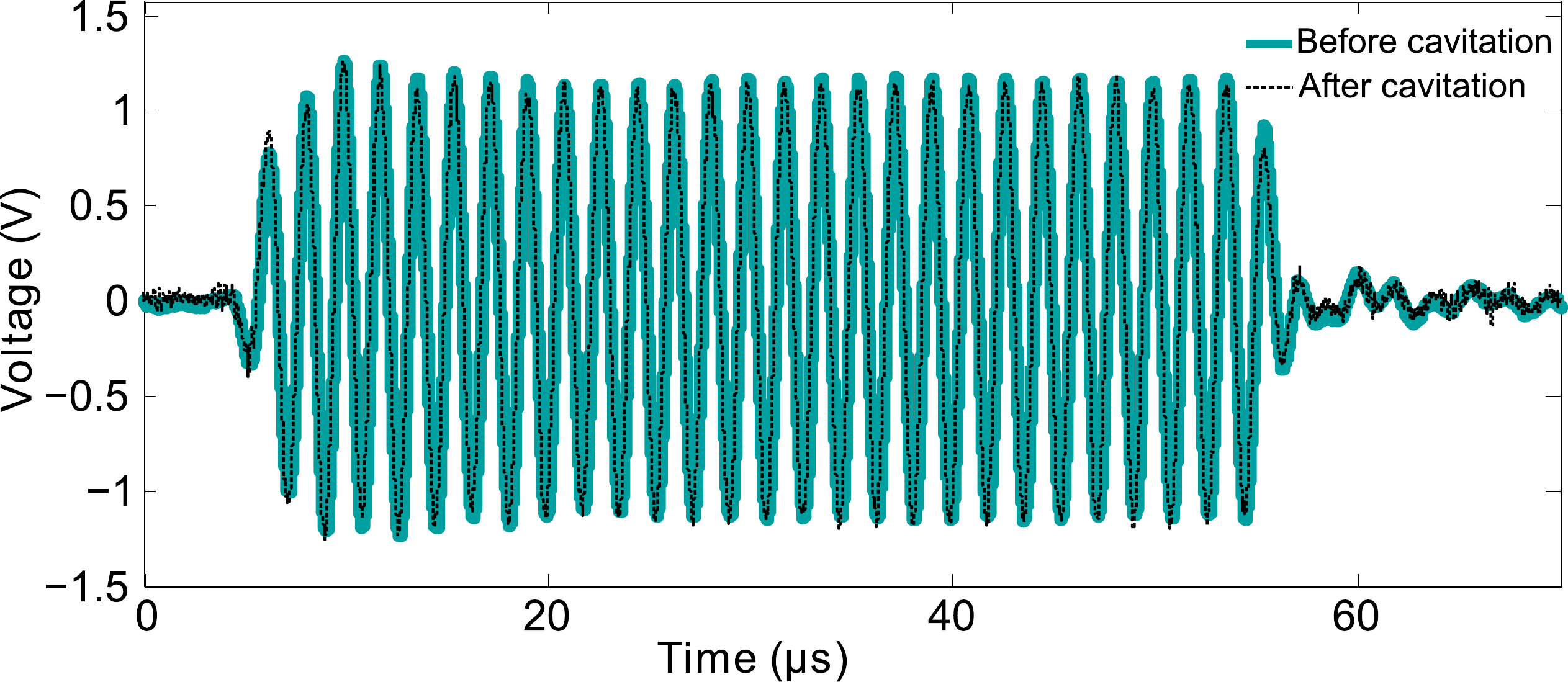}
	   \caption{\label{ResultatCavitation} Signal measured before (plain blue) and after (dotted black) cavitation. The signals are almost identical with a correlation coefficient equal to 0.9963, meaning that the cavitation had very little impact on the hydrophone's sensitivity.}
\end{figure}

The amplitude of the measured signals before and after 4800 seconds (80 minutes) of cavitation is plotted respectively in plain blue and in dashed black in figure (\ref{ResultatCavitation}).
The correlation coefficient $Q$ as defined in (\ref{coefCorrFormula}) is equal to 0.9963. Induced signal is consequently little impacted by the exposition to the cavitation cloud. Visual inspection of the wire showed only a small color change due to the destruction of the varnish.

\section{Discussion}
The objectives of the article are to characterize a Lorentz-force hydrophone prototype following IEC standard.

The hydrophone design is based on a tomographic approach \cite{grasland2012hydroEMproceeding}. The use of tomography is not uncommon for acoustic measurements, and it is used for Schlieren imaging \cite{pitts1994tomographic} and wire-based backscattering for transducer characterization \cite{raum1997}. If the acoustic field is symmetric, rotations are not necessary, which decreases significantly the number of required elementary measurements. In addition, the number of wires could easily be multiplied, potentially decreasing the number of acquisitions. However, the influence of each wire on its neighbours would have to be properly evaluated.

Due to the particular design of this hydrophone, the first step was to check that the magnetic field obtained had the required uniformity and that the mechanical tension exerted on the wire had no impact on the measurements. The magnetic field of a Halbach array proved to be very uniform, which facilitates a tomographic approach. The correlation between the measured signals for different tensions was larger than 0.96, and the wire tension does not seem to change the shape of the measured pulse pressures. For a wire 70 \micro m in diameter and 5 cm long, the wire’s mechanical resonance and first harmonic frequencies were under 100 Hz even for a 200 MPa  stress, far below medical ultrasound working frequencies \cite{grasland2013electromagnetic}. The experiment showed moreover that the mechanical tension had no significant effect on the measured signal. The wire can thus be considered as loose, following the fluid movement due to ultrasound. As such it is concluded that mechanical stress does not have any significant effect on the measurements, which considerably simplifies the fabrication and tuning of such hydrophones, and makes them very easy to manufacture compared with other hydrophone types.

As indicated in the IEC standard, the frequency response was studied. In the measured electric output characteristics, no peak of resonance was observed between 0.15 and 10 MHz, the first peak being located at 30 MHz. Electrical resonances consequently do not interfere with the Lorentz-force hydrophone in reception mode at the intended working frequencies. The upper cutoff frequency depends on wire diameter. The hydrodynamic model proposed is only valid when the wire diameter is small compared to ultrasound wavelength and does not predict the frequency bandwidth. This could not be explained by mechanical tension, given the mechanical resonance frequency as described in the previous paragraph. The upper cutoff frequency, around 2 MHz for 70 \micro m wire, is rather low for medical ultrasound calibration \cite{harris1988hydrophone}. Thinner wires could have higher cutoff frequencies, and lower cutoff frequencies should be studied, but it was not possible to explore these with the available equipment. Further studies could improve the model and help to increase the upper cut-off frequency, which is currently the main drawback of this prototype.

The directional response has also been evaluated. For most hydrophones, it is affected by the effective radius  of the active element, but the cylindrical shape of this Lorentz-force hydrophone makes this parameter irrelevant. For a rotation axis in the Y direction, the -3 dB directional response occurs at angles above $\pm35^o$ (limited by the PVC cylinder aperture); for a rotation axis in the X direction, the -3 dB response occurs at angles equal to $\pm25^o$.  However, the model of the directional response with the axis of rotation in the X direction supposes the ultrasound signal to be sinusoidal. If the signal is not sinusoidal, another directional response can be computed using the same equations.

The experimental sensitivity values must be compared with the ones given by the hydrodynamic model. Taking the density of water $\rho = 1000 \pm 10$ kg.m$^{-3}$, the speed of sound in water $c = 1470 \pm 10$ m.s$^{-1}$, the magnetic field on the wire $B = 0.17 \pm$ 0.04 T, the considered unit length $l = 1$ 1 mm, the correction factor $K = 5.0 \pm 0.3$ with $\rho_0 = 9000 \pm$ 500 kg.m$^{-3}$, the sensitivity without any amplification per unit length is equal to $2.3 \pm 0.7\ 10^{-11}$ V.mm$^{-1}$.Pa$^{-1}$, or -215 to -210 dB re \micro V.mm$^{-1}$.Pa$^{-1}$. This differs by 40\% from the experimental sensitivity. With a signal amplification $\alpha = 1\ 10^6$ V.A$^{-1}$ and an impedance of the wire plus amplifier of $R = 251 \pm 1 \Omega$, the sensitivity is equal to $9.4\pm 2.7\ 10^{-8}$ V.mm$^{-1}$.Pa$^{-1}$, or -143 to -138 dB re \micro V.mm$^{-1}$.Pa$^{-1}$, a difference of 10\% from the experimental sensitivity. The uncertainty in the magnetic field strength, about 25 \%, is the major uncertainty factor. These results show that the current hydrodynamic model gives a good representation of the phenomenon; however, calibration would still be needed for precise measurements.

Two problems occurr at low pressure, under 50 kPa: (1) the electromagnetic noise and the induced signal are of same order of magnitude and effective shielding is needed; and (2) the Reynolds number approaches 1 (for a 200 \micro m wire at 10 kPa, $Re = 1.3$) so that the main hypothesis for the hydrodynamic model is not fulfilled. Conversely, the signal should still be linear at pressures higher than 10 MPa. Moreover, the hydrophone has been shown to be highly resistant to cavitation. 

When comparing with other hydrophones, it should be borne in mind that a Lorentz-force hydrophone measures primarily fluid velocity, and that the given relationship between velocity and pressure is valid only for a plane wave. This hydrophone could however be combined with a pressure hydrophone to calculate the acoustic intensity without requiring the plane wave approximation, or to allow acoustic non linearities to be studied \cite{shaw2008buoyancy}.

If the hydrophone wire was to be destroyed, it would be easily replaceable, and at a very low cost. Damage to the wire varnish would have little impact on the signal, as the electrical conductivity of most electrolytes is small compared to wire conductivity (sea water conductivity = 5 S.m$^{-1}$, copper conductivity = 60 MS.m$^{-1}$) and current losses through the surrounding medium should be negligible.

\section{Conclusion}
This article presented some characteristics of a Lorentz-force hydrophone and evaluated the principal characteristics commonly specified for such devices. The current design and hydrodynamic model permits the high resolution 2D mapping of the acoustic velocity with a tomographic method. The tension of the wire was show to have very little impact on the measurements, which makes the Lorentz-force hydrophone the easier to fabricate. However the upper cut-off frequency response is between 1 and 2 MHz for 100 \micro m wire and is not explained by the current hydrodynamic model. The directional response was shown to depend on the axis of rotation but has a -3 dB response angle higher than $\pm25^o$ in both directions. The output signal is linear with pressure over a range from 50 kPa to 10 MPa and the hydrophone is highly resistant to cavitation. In addition to the low cost of the device, this hydrophone is particularly suitable for high pressure ultrasound medical applications such as ultrasonic lithotripsy, high intensity focused ultrasound, or industrial measurements in hostile environments.

\section*{Acknowledgments}
The authors would like to warmly thank Alexandre Vignot for modelling in 3D the hydrophone, Adrien Matias for helping on the mechanics, Alain Birer for helping on the electronic part, Frederic Padilla for providing the IEC standard and the language correction, Claude Inserra for theory suggestions on wire tension, Gail ter Haar and John Civale for the discussions about transducer metrology, and all the people who have checked the language.

\bibliographystyle{IEEEtran}
\bibliography{biblio}

\end{document}